\documentclass[prd,eqsecnum,showpacs,twocolumn,amsfonts,amssymb]{revtex4}

\usepackage{graphicx}

\usepackage{bm}

\newcommand{\beq}{\begin{equation}}
\newcommand{\eeq}{\end{equation}}
\newcommand{\beqs}{\begin{eqnarray}}
\newcommand{\eeqs}{\end{eqnarray}}

\newcommand{\gsim}{\mathrel{\raisebox{-
.6ex}{$\stackrel{\textstyle>}{\sim}$}}}

\newcommand{\drawsquare}[2]{\hbox{%
\rule{#2pt}{#1pt}\hskip-#2pt
\rule{#1pt}{#2pt}\hskip-#1pt
\rule[#1pt]{#1pt}{#2pt}}\rule[#1pt]{#2pt}{#2pt}\hskip-#2pt
\rule{#2pt}{#1pt}}
\newcommand{\fund}{\raisebox{-.5pt}{\drawsquare{6.5}{0.4}}}
\newcommand{\sym}{\raisebox{-.5pt}{\drawsquare{6.5}{0.4}}\hskip-0.4pt%
        \raisebox{-.5pt}{\drawsquare{6.5}{0.4}}}
\newcommand{\asym}{\raisebox{-3.5pt}{\drawsquare{6.5}{0.4}}\hskip-6.9pt%
        \raisebox{3pt}{\drawsquare{6.5}{0.4}}}

\begin{document}

\title{Study of an Alternate Mechanism for the Origin of Fermion Generations}

\author{Thomas A. Ryttov}

\author{Robert Shrock}

\affiliation{
C. N. Yang Institute for Theoretical Physics \\
Stony Brook University \\
Stony Brook, NY 11794}

\begin{abstract}

In usual extended technicolor (ETC) theories based on the group
${\rm{SU}(N_{ETC}})_{ETC}$, the quarks of charge 2/3 and $-1/3$ and the charged
leptons of all generations arise from ETC fermion multiplets transforming
according to the fundamental representation. Here we investigate a different
idea for the origin of SM fermion generations, in which quarks and charged
leptons of different generations arise from ETC fermions transforming according
to different representations of ${\rm{SU}(N_{ETC}})_{ETC}$.  Although this
mechanism would have the potential, {\it a priori}, to allow a reduction in the
value of $N_{ETC}$ relative to conventional ETC models, we show that, at least
in simple models, it is excluded by the fact that the technicolor sector is not
asymptotically free or by the appearance of fermions with exotic quantum
numbers which are not observed.

\end{abstract}

\pacs{12.60.-i,12.60.Nz,11.15.-q}

\maketitle

\section{Introduction} 

Three of the outstanding mysteries in particle physics at present are the
origin of electroweak symmetry breaking (EWSB) and the origin of the Standard
Model (SM) fermion generations and their associated hierarchy of masses.  It is
possible that electroweak symmetry breaking is dynamical, due to the formation
of a condensate of fermions subject to an asymptotically free vectorial gauge
interaction, technicolor (TC), that becomes strongly coupled at the TeV scale
\cite{tc}.  In order to give masses to the SM fermions, the technicolor theory
is embedded in a larger framework called extended technicolor (ETC) \cite{etc}.
Early studies modelled ETC effects via four-fermion operators added to the
technicolor Lagrangian.  Subsequently, reasonably ultraviolet-complete ETC
models were constructed, with detailed specification of the ETC field content
and symmetry breaking \cite{at94}-\cite{ckm}.  In these models the ETC gauge
symmetry is broken in a sequence of stages corresponding to the SM fermion
generations, leaving a subgroup which is the technicolor symmetry.  We take the
technicolor gauge group to be ${\rm SU}(N_{TC})_{TC} \subset {\rm
SU}(N_{ETC})_{ETC}$. The number of SM fermion generations (with associated
neutrinos lighter than $\sim m_Z/2$) is 3, but it will be useful to write this
in general as $N_{gen.}$ since it affects the structure of the ETC model.  In a
usual ETC theory of this sort the quarks of charge 2/3 and $-1/3$ and the
charged leptons arise from ETC fermion multiplets transforming according to the
fundamental representation of ${\rm SU}(N_{ETC})_{ETC}$, as a set of $N_{gen.}$
technicolor-singlet components corresponding to the SM generations, with the
remaining $N_{TC}$ components being technifermions. Hence,
$N_{ETC}=N_{gen.}+N_{TC}=3+N_{TC}$. Left-handed weak isodoublet neutrinos also
arise in this manner, while electroweak-singlet, right-handed neutrinos arise
as technicolor-singlet components of SM-singlet ETC multiplets, in such a
manner as to produce a requisite low-scale seesaw mechanism yielding the very
small observed neutrino masses \cite{nt}. In order to minimize technicolor
corrections to the $W$ and $Z$ propagators and to produce a technicolor theory
with a large but slowly running gauge coupling, a number of recent models have
used the smallest non-Abelian value of $N_{TC}$, namely $N_{TC}=2$.  There has
also been interest in technifermions in higher-dimensional representations of
the technicolor group \cite{higherrep}-\cite{sanrev}. Recent reviews of
theories with dynamical EWSB are given in Ref. \cite{sanrev}-\cite{dewsb}.

Because the origin of SM fermion generations is not understood, it is
worthwhile to explore various different possibilities for this origin.  For
example, studies assessing the feasibility of trying to embed theories with
dynamical EWSB in simple groups and to work out their implications for the
number of SM fermion generations have been carried out \cite{fs,tg}.  In
Ref. \cite{gen} we recently investigated the influence of the number of
generations, $N_{gen.}$, on the structure and breaking patterns of ETC theories
of the usual type, in which all SM-nonsinglet fermions arise as
technicolor-singlet components of fermions that transform as fundamental
representations of the ETC gauge group.

Here we investigate a different mechanism for the origin of SM fermion
generations, in which the SM fermions of a given type (quark of charge 2/3 or
$-1/3$ or lepton) belonging to different generations arise from ETC fermions
transforming according to different ETC representations.  This mechanism has
the potential appeal that, if it were feasible, it would allow one to construct
an ETC model with smaller values of $N_{ETC}$, namely $N_{ETC} <
N_{gen.}+N_{TC}$.  However, we shall show that, at least for the models that we
consider, it is not feasible.  Although our results are thus negative, we
believe that they yield useful insights into constraints on mechanisms to
explain SM fermion generations.

\section{An Illustrative Model}

We take the technicolor group to be ${\rm SU}(N_{TC})_{TC}$.  In order to
minimize technicolor corrections to the $W$ and $Z$ propagators, the value
$N_{TC}=2$ is favored, but we will show results for general $N_{TC}$. The
technifermions are taken to transform as SM families.  We shall show the
problems that one encounters with a simple illustrative model having only
$N_{gen.}=2$ SM fermion generations.  In this case, the minimal ETC
group has
\beq
N_{ETC}=N_{TC}+1 \ . 
\label{netc}
\eeq
As in previous ETC models \cite{at94,nt,ckm,gen}, we
will also use another auxiliary gauge interaction, denoted as hypercolor (HC),
with gauge group SU(2)$_{HC}$, that becomes strongly coupled above the TeV
scale.  The full gauge symmetry operative at scales $\mu \gsim 10$ TeV is thus
\beq
{\rm SU}(N_{ETC})_{ETC} \times {\rm SU}(2)_{HC} \times {\rm SU}(3)_c
\times {\rm SU}(2)_L \times {\rm U}(1)_Y
\label{gfull}
\eeq
As indicated in Eq. (\ref{gfull}), in this type of ETC model, the ETC gauge
bosons do not carry any SM quantum numbers.  The two generations of SM quarks
and leptons are arranged together with technifermions with the same SM quantum
numbers in ETC multiplets transforming according to the fundamental
representation, $F \equiv \fund$, and a rank-2 symmetric representation of
${\rm SU}(N_{ETC})_{ETC}$, $S \equiv \sym$. Note that
$d_{ETC,S}=N_{ETC}(N_{ETC}+1)/2$.  These fermions are
\begin{widetext}
\beq
Q^{a,i}_L: \ (N_{ETC},1,3,2)_{1/3,L} \ , \quad\quad
u^{a,i}_R: \ (N_{ETC},1,3,1)_{4/3,R} \ , \quad\quad
d^{a,i}_R: \ (N_{ETC},1,3,1)_{-2/3,R} \ ,
\label{quarks1}
\eeq
\beq
L^i_L: \ (N_{ETC},1,1,2)_{-1,L}, \quad\quad
e^i_R: \ (N_{ETC},1,1,1)_{-2,R} \ .
\label{leptons1}
\eeq
\beq
\tilde Q^{a,ij}_L: \ (d_{ETC,S},1,3,2)_{1/3,L} \ , \quad\quad
\tilde u^{a,ij}_R: \ (d_{ETC,S},1,3,1)_{4/3,R} \ , \quad\quad
\tilde d^{a,ij}_R: \ (d_{ETC,S},1,3,1)_{-2/3,R} \ ,
\label{quarks2}
\eeq
\beq
\tilde L^i_L: \ (d_{ETC,S},1,1,2)_{-1,L}, \quad\quad
\tilde e^i_R: \ (d_{ETC,S},1,1,1)_{-2,R} \ .
\label{leptons2}
\eeq
\end{widetext}
where $a$ and $i$ are SU(3)$_c$ and SU(3)$_{ETC}$ indices, respectively, the
numbers in parentheses are the dimensions of the representations of the
non-Abelian factor groups in Eq. (\ref{gfull}), and the subscripts denote the
weak hypercharge, $Y$. The fermions in Eqs. (\ref{quarks1})-(\ref{leptons2})
are assumed to have no mass terms in the high-scale Lagrangian invariant under
the group (\ref{gfull}), since if such masses were present, this would
explicitly violate the ${\rm SU}(2)_L \times {\rm U}(1)_Y$ electroweak
symmetry.  Since this is just an illustrative model, it is accepted that the
beta function for color SU(3)$_c$ would differ from its form in the real world.
Above the EWSB scale, color SU(3)$_c$ would actually not be asymptotically
free; the leading coefficient of its beta function (see appendix for notation)
would be $(b_1)_{{\rm SU}(3)_c} = -1$. If the technicolor interaction worked
coras usual, producing technifermion condensates that break electroweak
symmetry, then below the EWSB scale, SU(3)$_c$ would become asymptotically
free, with two generations of quarks, and hence $(b_1)_{{\rm SU}(3)_c} = 25/3$.

The model also contains a set of SM-singlet fermions, 
\beq
\psi_{i,R} \ : \quad (\overline{N}_{ETC},1,1,1)_0 \ , 
\label{psi}
\eeq
\beq
\chi^{ij}_R \ : \quad (d_{ETC,S},1,1,1)_0 \ , 
\label{chi}
\eeq
\beq
\zeta^{i \alpha}_R \ : \quad (N_{ETC},2,1,1)_0 \ , 
\label{zeta}
\eeq
and
\beq
\omega_{\alpha,p,R} \ : \quad (1,2,1,1)_0  \ . 
\label{omega}
\eeq
If $N_{ETC}=N_{TC}+1$ is odd (even), then the number of copies of
$\omega_{\alpha,p,R}$ is one (two); the copy index is denoted $p$.  Hence, the
total number of chiral SU(2)$_{HC}$ doublets is even, and the HC sector is free
of a global Witten anomaly associated with $\pi_4({\rm SU}(2))={\mathbb
Z}_2$. This ETC theory is a chiral gauge theory with no anomalies in gauged
currents.  The fermions in Eqs. (\ref{psi})-(\ref{omega}) are taken to have
zero mass terms in the high-scale Lagrangian, consistent with the chiral gauge
symmetry.  The model is constructed so that the TC-singlet components of 
$\psi_R$ and $\chi_R$, namely $\psi_{1,R}$ and $\chi^{11}_R$, could form
right-handed SM neutrino states.

Before proceeding with the analysis, we remark briefly on some properties that
the model would exhibit if the technicolor sector were asymptotically free and
hence produced condensates of technifermions. Studies of the Dyson-Schwinger
equation for a technifermion with zero Lagrangian mass, transforming according
to the representation $R$ of the technicolor group, show that, in the
one-technigluon exchange approximation, if $\alpha_{TC}(\mu)C_2(R) \gsim O(1)$
(where $C_2(R)$ is the quadratic Casimir invariant for this representation),
then there is a solution with a nonzero, dynamically generated technifermion
mass \cite{sd}. This reflects the formation of a corresponding bilinear
technifermion condensate.  Since the technicolor sector of the present model
contains technifermions transforming under two different representations, the
fundamental, $F$, and the symmetric rank-2 tensor, $S$, it follows that, if the
technicolor theory were asymptotically free so that $\alpha_{TC}(\mu)$
increased with decreasing scale $\mu$, then the first condensation would
involve the technifermions in the larger representation, $S$, and would occur
when
\beq
 \alpha_{TC}(\mu) C_2(S_{{\rm SU}(N_{TC})}) \sim O(1) 
\label{alfcritsym}
\eeq
where
\beq
C_2(S_{{\rm SU}(N_{TC})}) = \frac{N_{TC}+2)(N_{TC}-1)}{N_{TC}} \ . 
\label{c2sym}
\eeq
The associated scale is denoted as $\Lambda_{TC,S}$ and the corresponding
technicolor analogue of the pion decay constant as $f_{TC,S}$.  At a lower
scale, $\Lambda_{TC,F}$, where 
\beq
 \alpha_{TC}(\mu) C_2(F_{{\rm SU}(N_{TC})}) \sim O(1) 
\label{alfcritfund}
\eeq
where $C_2(F_{{\rm SU}(N_{TC})}) = (N_{TC}^2-1)/(2N_{TC})$, the
technifermions in the fundamental representation of the technicolor group would
condense.  The associated parameter $f_{TC,F}$ would satisfy 
\beq
\frac{f_{TC,F}}{f_{TC,S}} = \frac{\Lambda_{TC,F}}{\Lambda_{TC,S}} \ .  
\eeq
The technifermion condensation at the higher scale, $\Lambda_{TC,S}$, would
play the dominant role in breaking electroweak symmetry and giving the $W$ and
$Z$ masses.  To the extent that $f_{TC,F}<<f_{TC,S}$, these would have
the approximate form 
\beq
m_W^2 \simeq \frac{g^2 N_{TD} (f_{TC,S})^2}{4} 
\label{mwsq}
\eeq
and
\beq
m_Z^2 \simeq \frac{(g^2+g'^2) N_{TD} (f_{TC,S})^2}{4} \ , 
\label{mzsq}
\eeq
where $g$ and $g'$ are the gauge couplings for SU(2)$_L$ and U(1)$_Y$ and 
$N_{TD}=N_c+1=4$ is the number of technifermion SU(2)$_L$ doublets arising
from the $S$ representation of ${\rm SU}(N_{TC})_{TC}$.

The SM fermions arising from the TC-singlet components of the $F$
representations of the ETC group would get masses via diagrams in which they
would emit a virtual massive ETC vector boson $V^1_j$ with $2 \le j \le
N_{TC}$, transforming to a technifermion and reabsorbing this boson.  For
example, there would be a diagram in which a $e^1_R$ would transform in this
way to a $e^j_R$ technilepton, which, after a dynamical mass insertion $\sim
\Lambda_{TC,F}$, would change to a $e^j_L$, reabsorb the virtual $V^1_j$, and
become a final-state $e^1_L$, producing the electron mass term $m_e \bar e_L
e_R$.  In contrast, the SM fermions arising from the TC-singlet components of
the $S$ representation would get two types of contributions to their masses.
The first of these would be similar to the usual type of diagram described
above, in which, for example, a $\tilde e^{11}_R$ would emit a virtual $V^1_j$
with $2 \le j \le N_{TC}$, going to the technifermion $\tilde e^{1j}_R$, which,
after a mass insertion $\sim \Lambda_{TC,F}$, would change to a $e^{1j}_L$,
reabsorbing the virtual $V^1_j$ and becoming a final-state $e^{11}_L$. This
would give a contribution to $m_{\tilde e} \bar{\tilde e}_L \tilde e_R$ similar
to $m_e \bar e_L e_R$.  The second type of mass generation for the SM fermions
arising from the $S$ ETC multiplets would involve two-loop diagrams such as one
in which an $\tilde e^{11}_R$ would emit a virtual $V^1_j$, going to a
$e^{1j}_R$, which would emit a virtual $V^1_k$, with $2 \le j,k \le N_{TC}$
going to a $e^{kj}_R$. After the larger technifermion mass insertion
$\sim \Lambda_{TC,S}$, the $e^{kj}_R$ would become an $e^{kj}_L$.  This would
reabsorb the virtual $V^1_k$ becoming a $e^{1j}_L$, which would then reabsorb
the $V^1_j$, becoming the final-state $\tilde e^{11}_L$.  This diagram would
thus give a different contribution to the mass term $m_{\tilde e}
\bar{\tilde{e}}_L \tilde{e}_R$ so that $m_e$ would differ from $m_{\tilde e}$.
(Other diagrams would also contribute.)  At first sight, this mechanism would
thus appear to have the possibility to produce various masses for SM fermions
in a manner different from that of usual ETC models and, if this mechanism were
feasible in such models, then a detailed analysis of the two-loop contributions
would be in order. Additional ingredients would be needed in order to account
for full intergenerational mixing between quarks (Cabibbo-Kobayashi-Maskawa
mixing) and the corresponding mixing between leptons of different generations,
because ETC transitions by themselves would not mix different ETC
representations.  We next proceed with the analysis of the ETC sector.

\section{Calculations}

The theory must contain a plausible mechanism to break the ETC gauge symmetry
at down to the residual exact TC symmetry.  The ETC breaking at scales above
the electroweak scale must avoid EWSB, which should occur at the electroweak
scale via formation of the technifermion condensates.  In a conventional ETC
theory in which the quarks and charged leptons arise as TC-singlet
components of fermions transforming as fundamental representations of the ETC
group, one is cognizant of the possible condensation channel
\beq
F \times \bar F \to 1
\label{ffbar}
\eeq
involving these SM-nonsinglet fermions.  This condensation channel must be
avoided for a number of reasons: (i) it would break electroweak symmetry at too
high a scale; (ii) it would not break the ETC gauge symmetry, and hence (iii)
it would not separate the usual SM fermions from the technifermions.  An
approximate measure of the attractiveness of a channel $R_1 \times R_2 \to
R_{cond.} $ is
\beq
\Delta C_2 = C_2(R_1)+C_2(R_2)-C_2(R_{cond.}) \ , 
\label{deltac2}
\eeq
where $R_j$ denotes the representation under a relevant gauge interaction and
$C_2(R)$ is the quadratic Casimir invariant for the representation $R$.  For
the channel (\ref{ffbar}) this is
\beq
\Delta C_2 = 2C_2(F) = \frac{N_{_{ETC}}^2-1}{N_{_{ETC}}} \ . 
\label{c2ffbar}
\eeq
The situation is made more difficult by the presence of SM-nonsinglet fermions
transforming in higher-dimensional ETC representations.  For the $S$
representation of ${\rm SU}(N_{ETC})_{ETC}$, there is the possible condensation
channel
\beq
S \times \bar S \to 1
\label{fsymfsymbar}
\eeq
for which the measure of attractiveness is 
\beq
\Delta C_2 = 2C_2(S) = \frac{2(N_{_{ETC}}+2)(N_{_{ETC}}-1)}{N_{_{ETC}}} \ . 
\label{c2symsymbar}
\eeq

In quasi-realistic ETC models with $N_{gen.}=3$ and ETC-nonsinglet
SM-nonsinglet fermions only in the fundamental representation of the ETC gauge
group \cite{at94,nt,ckm}, so that $N_{ETC}=N_{gen.}+N_{TC}=5$, one avoids the
occurrence of the unwanted condensation in the channel (\ref{ffbar}) by a
hybrid mechanism that makes use of the fact that the models contain
ETC-nonsinglet, SM-singlet chiral fermions transforming as the antisymmetric
rank-2 (10-dimensional) tensor representation of SU(5)$_{ETC}$, and these can
condense in a first stage of ETC symmetry breaking with a $\Delta C_2 = 24/5$
that is equal to the $\Delta C_2$ for the unwanted channel (\ref{ffbar}).  The
second and third stages of ETC symmetry breaking are caused primarily by the
auxiliary hypercolor gauge interaction, which becomes strongly coupled above
the TeV level. In the present model we will also use the HC interaction to
produce the desired breaking of ETC to TC symmetry.  However, this does mean
that one needs to assume a substantially stronger HC coupling $\alpha_{_{HC}}
>> \alpha_{_{ETC}}$ at scales above a TeV, so as to produce the necessary
breaking of ETC symmetry and to avoid the undesired high-scale condensation of
ETC fermions in either of the channels (\ref{ffbar}) and (\ref{fsymfsymbar}).

The leading coefficient of the HC beta function is 
\beq
(b_1)_{HC} = \frac{1}{3}\left [ 20-(N_{TC}+\Delta n_\omega) \right ] \ , 
\label{b1hc}
\eeq
where $\Delta n_\omega=0$ for even $N_{TC}$ and 1 for odd $N_{TC}$.  Since we
envision $N_{TC}=2$ or 3, the HC interaction is asymptotically free, and its
coupling grows as the energy scale decreases.  The size of this coupling at
high scales is taken to be large enough so that, at a scale above 1 TeV that we
shall denote $\Lambda_1$, it becomes sufficiently strong to form the bilinear
condensation in the $2 \times 2 \to 1$ channel, with associated condensate
\beq
\langle \zeta^{i \alpha \ T}_R C \omega_{\alpha,p,R} \rangle \ . 
\label{zetaomega_condensate}
\eeq
This breaks the ETC symmetry ${\rm SU}(N_{TC}+1)_{ETC}$ to ${\rm
SU}(N_{TC})_{TC}$.  With no loss of generality, we take the breaking direction
in the ETC group space to be given by $i=1$ (so that $2 \le i \le N_{TC}$ are
TC indices).  The $\zeta^{1 \alpha}_R$ and $\omega_{\alpha,p,R}$ fermions
involved in the condensate (\ref{zetaomega_condensate}) get dynamical masses of
order $\Lambda_1$.  Other condensation channels depend on the value of
$N_{TC}$.  For example, for the favored, minimal case, $N_{TC}=2$ (for which 
there is only a single $\omega_{\alpha,R}$ field), the HC interaction would
also produce a condensate of the form
\beq
\langle \epsilon_{1jk} \epsilon_{\alpha\beta} \, 
\zeta^{j \alpha \ T}_R C \zeta^{k \beta}_R \rangle 
\label{zetazeta_condensate}
\eeq
which is invariant under both SU(2)$_{TC}$ and SU(2)$_{HC}$.  This condensate
would give a dynamical mass of order $\Lambda_1$ to the remaining SM-singlet,
HC-nonsinglet fermions in the model, $\zeta^{j,\alpha}_R$ with $j=2,3$ and
$\alpha=1,2$.  Thus, for $N_{TC}=2$, all SM-singlet, HC-nonsinglet fermions
would be integrated out of the effective field theory at mass scales below
$\Lambda_1$.  For $N_{TC} \ge 3$, the condensate (\ref{zetaomega_condensate})
would form in the same way, and the details of other condensates formed will
not be important for our analysis below.

The $S = \sym$ representation of ${\rm SU}(N_{ETC})$ decomposes with
respect to the subgroup ${\rm SU}(N_{TC})$ as
\beq
\sym_{{\rm SU}(N_{ETC})} = \{ 1 + \fund + \sym \}_{{\rm SU}(N_{TC})} \ . 
\label{symdecomp}
\eeq
Note that the $\asym_{{\rm SU}(N_{ETC})}$ has a corresponding decomposition 
\beq
\asym_{{\rm SU}(N_{ETC})} = \{ \fund + \asym \}_{{\rm SU}(N_{TC})}
\label{asymdecomp}
\eeq
that does not contain any TC-singlet component, so we could not use this as
a source of another SM fermion in the way that we have with the $\sym$
representation.

Having discussed how the ETC theory breaks, we examine the resultant
technicolor theory and find a serious problem, namely that it is not
asymptotically free. We calculate that the contribution to the leading
coefficient of the technicolor beta function from the SM-nonsinglet
technifermions plus the right-handed technineutrino, by themselves, is enough
to render the theory non-asymptotically free:
\beq
(b_1)_{TC} = -\frac{1}{3}(5N_{TC}+64) \ . 
\label{b1tc}
\eeq
For the minimal case, $N_{TC}=2$, this is the full $b_1$, while for $N_{TC} \ge
3$, depending on the type of ETC breaking, there could be additional
contributions from other SM-singlet technifermions.  However, since these would
just make this coefficient more negative, we do not have deal with them in
detail.  The lack of asymptotic freedom in the technicolor sector excludes this
mechanism for obtaining SM generations because it means that the technicolor
theory does not confine and does not produce the technifermion condensates that
are the source of dynamical electroweak symmetry breaking.  In addition to our
simple model with $N_{gen.} =2$, we have also investigated models with larger
numbers of generations and larger ETC gauge groups, and we find that the loss
of asymptotic freedom in the technicolor sector appears to be a generic problem
with these one-family models in which one attempts to get SM fermions of
different generations from different types of technicolor representations.
It may be noted that the ETC theory itself is also not asymptotically free,
although it does not have to be, in view of the fact that the ETC symmetry is 
broken to the technicolor subgroup by the hypercolor interaction.  We calculate
\beq
(b_1)_{ETC} = -\frac{5}{3}(N_{ETC}+10) = -\frac{5}{3}(N_{TC}+11) \ . 
\label{b1etc}
\eeq

\section{A TC/ETC Model with $[G_{ETC},G_{SM}] \ne 0$} 

A different type of technicolor model features technifermions that form
a left-handed color-singlet SU(2)$_L$-doublet with $Y=0$ and corresponding
right-handed fields with $Y=\pm 1$: 
\beq
{\xi^i \choose \eta^i}_L \quad \xi^i_R, \quad \eta^i_R 
\label{1dtc}
\eeq
where $1 \le i \le N_{TC}$.  Thus, $\xi$ and $\eta$ have electric charges $\pm
1/2$. Although this TC sector is simpler than conventional one-family TC
models, the embedding in ETC is more complicated, since the ETC gauge group,
$G_{ETC}$, does not commute with the SM gauge group, $G_{SM}$, so the ETC gauge
bosons carry SM quantum numbers.  One approach to the embedding has been to
place the SM fermions in vectorlike fundamental representations, so that the
full ETC gauge group is ${\rm SU}(N_{ETC})_{ETC}$ with
$N_{ETC}=N_{TC}+N_{gen.}(N_c+1)$. Again, the TC gauge group ${\rm
SU}(N_{TC})_{TC}$ is a subgroup of this ETC group. However, the breaking of the
large ETC symmetry down to the TC gauge symmetry is considerably more
complicated than the breaking of the one-family ETC theory with its much
smaller value of $N_{ETC}=N_{gen.}+N_{TC}$. 

In the present context, one might envision constructing a model of this sort 
with one set of technifermions transforming as the $F$ representation of ${\rm
SU}(N_{TC})_{TC}$, as in Eq. (\ref{1dtc}), and another transforming as the $S$
representation, 
\beq
{\tilde \xi^{ij} \choose \tilde \eta^{ij}}_L \quad \tilde \xi^{ij}_R, 
\quad \tilde \eta^{ij}_R \ . 
\label{1dtcsym}
\eeq
With these technifermions by themselves, the TC theory is asymptotically free; 
\beq
(b_1)_{TC} = \frac{1}{3}(7N_{TC}-12) \ . 
\label{b1tc1d}
\eeq

One would then combine some subset of SM fermions with the
technifermions in vectorlike $F$ representations of the ETC group, 
and the orthogonal subset of SM fermions with the technifermions in the $S$ 
representation of the ETC group.  For example, if one combined the first
generation of SM fermions with technifermions in an ETC multiplet transforming
as the $F$ representation, one would have 
\beq
F_{L,R} = \left ( \begin{array}{ccc}
       \{ \xi^i \} & \{u^a\} & \nu_e \\
      \{ \eta^i \} & \{d^a\} & e \end{array} \right )_{L,R}\
\label{fmult}
\eeq
where $1 \le i \le N_{TC}$, $1 \le a \le N_c$, $\{ \xi^i \} \equiv
\{\xi^1,...,\xi^{N_{TC}}\}$, $\{u^a \} \equiv \{u^1,u^2,u^3\}$, and so forth
for the other $\{...\}$ sets.  If this embedding worked, then again one could
achieve a reduction of $N_{ETC}$; here, it would be $N_{ETC}=N_{TC}+N_c+1$.
But the embedding of the $S$ technifermions in an ETC multiplet is problematic.
As was pointed out in Ref. \cite{ts}, because of the fact that
$[G_{ETC},G_{SM}] \ne 0$, a higher-representation ETC multiplet contains
techni-singlet fermions with exotic quantum numbers which are not observed.
For example, these include color $\underline{6}$ fermions, leptoquarks,
fermions with charge $-2$ and lepton number $L=2$, etc.  This excludes a model
with technifermions (\ref{1dtc}) and (\ref{1dtcsym}) in both $F$ and $S$
representations.

In summary, we have studied the feasibility of an alternate mechanism for
explaining Standard-Model fermion generations in the context of models with
dynamical electroweak symmetry breaking.  In this mechanism, quarks and charged
leptons of different generations would arise from ETC fermions transforming
according to different representations of the ETC gauge group.  We have shown
that in models in which technifermions transform as SM families, this would
render the TC sector non-asymptotically free, and in TC models with
technifermions of the type (\ref{1dtc}) and (\ref{1dtcsym}), it would lead to
unobserved fermions with exotic quantum numbers.  Although these results are
negative, we believe that they are useful, since they show the restrictions on
how one includes generations in ETC theories.

This research was partially supported by the grant NSF-PHY-06-53342.

\section{Appendix} 

Here we define some notation used in the text. For a gauge group
$G_j$ we denote the running gauge coupling as $g_j(\mu)$, where $\mu$ is the
Euclidean reference momentum, and we denote
$\alpha_j(\mu)=g_j(\mu)^2/(4\pi)$. The beta function is $\beta_{G_j} =
dg_j/dt$, where $dt = d\ln \mu$.  We write
\beq
\frac{d\alpha_j}{dt} = - \frac{\alpha_j^2}{2\pi} \left [ b_1 + 
\frac{b_2 \, \alpha_j}{4\pi} + 
O(\alpha_j^3) \right ] 
\label{beta}
\eeq
where the first two coefficients, $b_1$ and $b_2$, are scheme-independent.  
For a representation $R$ of a Lie group $G$, the quadratic Casimir
invariant $C_2(R)$ is defined by 
$\sum_{a=1}^{order(G)} \sum_{j=1}^{dim(R)} (T_a)_{ij}(T_a)_{jk} = 
C_2(R)\delta_{ik}$.


\begin{thebibliography}{99}

\bibitem{tc} 
S. Weinberg, Phys. Rev. D {\bf 19} (1979) 1277;
L. Susskind, Phys. Rev. D {\bf 20} (1979) 2619.

\bibitem{etc}
S. Dimopoulos and L. Susskind, Nucl. Phys. B {\bf 155} (1979) 237;
E. Eichten and K. Lane, Phys.  Lett. B {\bf 90} (1980) 125.

\bibitem{at94}
T. Appelquist and J. Terning, Phys. Rev. D {\bf 50} (1994) 2116. 

\bibitem{nt}
T. Appelquist and R. Shrock, Phys. Lett. B {\bf 548} (2002) 204; 
Phys. Rev. Lett. {\bf 90} (2003) 201901. 

\bibitem{ckm}
T. Appelquist, M. Piai, and R. Shrock, Phys. Rev. D {\bf 69} (2004) 015002. 

\bibitem{higherrep}
K. Lane and E. Eichten, Phys. Lett. B {\bf 222} (1989) 274. 

\bibitem{sann12}
D. Hong, S. Hsu, and F. Sannino, Phys. Lett. B {\bf 597} (2004) 89;
F. Sannino and K. Tuominen, Phys. Rev. D {\bf 71} (2005) 051901(R).

\bibitem{ts}
N. D. Christensen and R. Shrock, Phys. Lett. B {\bf 632} (2006) 92. 

\bibitem{hr2}
D. D. Dietrich and F. Sannino, Phys. Rev. D {\bf 75} (2007) 085018;
S. B. Gudnason, T. A. Ryttov, and F. Sannino, Phys. Rev. D {\bf 76}
(2007) 015005;
R. Foadi, M. T. Frandsen, T. A. Ryttov, and F. Sannino, Phys. Rev. D 
{\bf 76} (2007) 055005. 

\bibitem{sanrev}
F. Sannino, ArXiv:0911.0931. 

\bibitem{nag06}
R. Shrock, in {\it Proc. Internat. Workshop on the Origin of
Mass and Strongly Coupled Gauge Theories - SCGT06}, eds. M. Harada, M.
Tanabashi, and K. Yamawaki (World Scientific, Singapore, 2008), p. 227
(hep-ph/0703050).

\bibitem{sekhar}
R. S. Chivukula, M. Narain, and J. Womersley, Phys. Lett. B {\bf 667}
(2008) 1258 and http://pdg.lbl.gov.

\bibitem{dewsb}
See talks in {\it Workshop on Dynamical Electroweak Symmetry Breaking},
Southern Denmark University, 2008, http://hep.sdu.dk/dewsb.

\bibitem{fs}
E. Farhi and L. Susskind, Phys. Rev. D {\bf 20} (1979) 3404.

\bibitem{tg}
N. D. Christensen and R. Shrock, Phys. Rev. D {\bf 72} (2005) 035013; 
N. Chen and R. Shrock, Phys. Rev. D {\bf 78} (2008) 035002. 

\bibitem{gen}
T. Ryttov and R. Shrock, ArXiv:1004.2075.

\bibitem{sd}
%
T. Appelquist, K. Lane, and U. Mahanta, Phys. Rev. Lett. {\bf 61} (1988) 1553;
T. Appelquist, J. Terning, and L. C. R. Wijewardhana,
Phys. Rev. Lett.  {\bf 77} (1996) 1214. 

\end{thebibliography}
\end{document}